# Observation of the out-of-plane polarized spin current from CVD grown WTe$_2$

*Shuyuan Shi[1], Jie Li[2], Chuang-Han Hsu[1,3], Kyusup Lee[1], Yi Wang[1,4], Li Yang[2], Junyong Wang[5,6], Qisheng Wang[1], Hao Wu[2], Wenfeng Zhang[2], Goki Eda[5,6], Gengchiau Liang[1], Haixin Chang[2]\*, and Hyunsoo Yang[1,5]\**

[1]Department of Electrical and Computer Engineering, National University of Singapore, 117576, Singapore

[2]Quantum-Nano Matter and Device Lab, State Key Laboratory of Material Processing and Die & Mould Technology, School of Materials Science and Engineering, Huazhong University of Science and Technology, Wuhan 430074, China

[3]Institute of Physics, Academia Sinica, Taipei 11529, Taiwan

[4]Key Laboratory of Materials Modification by Laser, Ion and Electron Beams (Ministry of Education), Dalian University of Technology, Dalian 116024, China

[5]Centre for Advanced 2D Materials, National University of Singapore 6 Science Drive 2, 117546, Singapore

[6]Department of Physics, National University of Singapore, 117542, Singapore

**Weyl semimetal Td-phase WTe$_2$ possesses the spin-resolved band structure with strong spin-orbit coupling, holding promises as a useful spin source material. The noncentrosymmetric crystalline structure of Td-WTe$_2$ endows the generation of the out-of-plane polarized spin, which is of great interest in magnetic memory applications. Previously, WTe$_2$ was explored in spin devices based on mechanically exfoliated single crystal flakes with a size of micrometers. For practical spintronics applications, it is highly desirable to implement wafer-scale thin films. In this work, we utilize centimeter-scale chemical vapor deposition (CVD)-grown Td-WTe$_2$ thin films and study the spin current generation by the spin torque ferromagnetic resonance technique. We find the in-plane and out-of-plane spin conductivities of $7.36 \times 10^3$ ($\hbar/2e$) $(\Omega m)^{-1}$ and $1.76 \times 10^3$ ($\hbar/2e$) $(\Omega m)^{-1}$, respectively, in CVD-growth 5 nm-WTe$_2$. We further demonstrate the current induced magnetization switching in WTe$_2$/NiFe at room temperature in the domain wall motion regime, which may invigorate potential spintronic device innovations based on Weyl semimetals.**

# 1. Introduction

Highly efficient charge-spin conversion is on demand for the spintronic device applications with ultra-low power consumption. Over the past decade, the spin current generation in heavy metals has been widely explored for this endeavor.[1-2] With the emergence of topological quantum matters, much more efficient charge-spin conversion processes were discovered in topological materials[3-5] and two-dimensional interfaces[6-8] due to their peculiar electronic band structures. Recently, a new class of quantum matters, Weyl semimetal,[9-14] has sparked intense research interests in spintronics due to its non-trivial topological band structures that are expected to contribute significant spin conductivities in the bulk.[15] Among them, the transition metal dichalcogenide (TMD) based type-II Weyl semimetal,[9, 14] Td-phase $WTe_2$, has been experimentally shown to possess a spin-rich textured Fermi surface[16-18] and theoretically predicted to have decent spin conductivities due to the strong spin-orbit coupling.[19] Furthermore, conduction electrons on its bulk and surface states possess the characteristic of the spin-momentum-locking.[20] In addition, due to the low-symmetry crystal structures of TMD Weyl semimetals, the out-of-plane polarized spin currents were observed,[21-23] which makes it possible to switch the magnetization with perpendicular anisotropy without an external assist magnetic field. Recently, efficient charge-spin conversion processes in exfoliated $WTe_2$ flakes were reported.[23-25] The magnetization switching using the spin current from exfoliated Weyl semimetals $WTe_2$ and $MoTe_2$ flakes were further demonstrated.[23, 26] With above novel discoveries, TMD Weyl semimetals have been revealed as a new contender for spintronic device innovations.

Previously explored $WTe_2$ devices were mainly prepared by mechanical exfoliation from $WTe_2$ bulk crystals and thus normally in a form of micro-sized flakes located randomly on the wafer.[21-24, 27] Even though mechanical exfoliation ensures a good crystallinity of the material and facilitates the fundamental research, it is impractical for applications which require wafer-scale thin films. Therefore, quantification and understanding of the spin generation properties in large-scale $WTe_2$ thin films are of importance for applications and development.

In this study, we investigate the spin current generation in centimeter-scale, chemical vapor deposition (CVD)-grown $WTe_2$ thin films by performing spin torque ferromagnetic resonance (ST-FMR) measurements. We evaluate the in-plane and out-of-plane spin conductivities in CVD-$WTe_2$ thin films at room tempareture and compare with those from $WTe_2$ single crystal flakes. The calculation based on the first-principles electronic structures can account for the observed spin conductivites. Moreover, we demonstrate the magnetization switching of the in-plane magnetizated NiFe bar using the spin current generated by $WTe_2$ with a current density $J_c$ of $2.53 \times 10^5$ A cm$^2$.

## 2. Results and Discussion

Centimeter-scale $WTe_2$ thin films were synthesized on Si/$SiO_2$ wafers inside a three-zone CVD system by tuning the precursors and vapor deposition conditions.[29] **Figure 1**a presents the optical microscopy image of a typical CVD grown $WTe_2$ thin film uniformly covering the Si/$SiO_2$ substrate larger than 1 cm$^2$. The Raman spectra in Figure 1b shows the characteristic resonance modes $A_1^2$, $A_2^2$, $A_2^4$, $A_1^3$, $A_1^4$, $A_1^7$, and $A_1^9$ located at 80.6, 89.3, 110.5, 114.9, 132.0, 161.4, and 209.1 cm$^{-1}$, in agreement with that of $WTe_2$ thin flakes exfoliated from $WTe_2$ single crystals, indicating the Td phase of $WTe_2$.[30] The Raman mapping of the $A_1^7$ peak for the thin film shown in the inset of Figure 1b confirms a uniform coverage of Td-$WTe_2$ film on the Si/$SiO_2$ wafer. The $WTe_2$ thin film has ~7 monolayers with a thickness of ~5.0 nm measured by atomic force microscopy (AFM) (Figure 1c). The X-ray photoelectron spectroscopy (XPS) data in Figure 1d show the chemical states of W 4f and Te 3d electrons. The binding energy of W $4f_{5/2}$ (33.6 eV), W $4f_{7/2}$ (31.5 eV), Te $3d_{3/2}$ (583.3 eV) and Te $3d_{5/2}$ (572.9 eV) reflects the valence states of W (+4) and Te (-2) in the WTe$_2$ film. The X-ray diffraction (XRD) data of WTe$_2$ thin film in Figure 1e show a peak of crystal planes (002n, n=1), indicating that the film stack is along the c-axis of crystal, which is typical for ultrathin WTe$_2$ crystals.[31]

In order to quantify both the in-plane and out-of-plane polarized spin currents generated from the WTe$_2$ spin source layer, we utilize the ST-FMR technique.[21, 28, 32] **Figure 2**a depicts the schematic diagram of the ST-FMR set-up with the film structure, which consists of WTe$_2$ (5 nm)/NiFe (8 nm)/SiO$_2$ (5 nm). The NiFe and SiO$_2$ layers were deposited on WTe$_2$ using magnetron sputtering at room temperature with a base pressure < $2\times10^{-9}$ Torr. The fabrication processes of ST-FMR devices are detailed in experimental section. As depicted in the left panel of Figure 2a, an in-plane radio frequency (rf) current $I_{rf}$ with frequencies (f) ranging from 6 to 9 GHz and a power of 15 dBm is applied across the WTe$_2$/NiFe bilayer along the x-axis using a signal generator. Note that the current injection direction is close to the a-axis of WTe$_2$ as evidenced from the THz measurements (Supporting information S6). An external magnetic field B is swept in plane with an angle ($\theta_B$) of 40° with respect to the x-axis in order to satisfy the ferromagnetic resonance condition. As a consequence of $I_{rf}$, oscillating spin currents are generated in WTe$_2$ and diffuse into the NiFe layer, thereby exerting oscillating spin-orbit torques (SOTs) on NiFe magnetic moments including both the damping-like SOT (**m** × **σ** × **m**) and the field-like SOT (**σ** × **m**), where **m** and **σ** are the vectors of the magnetization in NiFe and the induced spin from WTe$_2$, respectively. $I_{rf}$ also exerts the Oersted field torque (**m** × **$H_{Oe}$**) on NiFe. These combined torques, which can be decomposed to the in-plane oriented torque $\tau_\parallel$ and out-of-plane oriented torque $\tau_\perp$ schematically shown in the left panel of Figure 2a, drive the NiFe magnetization away from equilibrium and into precession, and yields the time-dependent change of the anisotropic magnetoresistance of NiFe. Consequently, the change of the device resistance mixing with $I_{rf}$ gives rise to a d.c. voltage which is measured as the ST-FMR signal $V_{mix}$ by a lock-in amplifier. The circuit for detecting $V_{mix}$ is schematically shown in the right panel of Figure 2a.

Figure 2b shows $V_{mix}$ as a function of magnetic field $B$ obtained from a WTe$_2$ (5 nm)/NiFe (8 nm) device. The inset of Figure 2b shows $f$ as a function of resonance magnetic field $B_0$, which can be fitted by the Kittel formula $f = \gamma/2\pi[B_0(B_0 + \mu_0 M_{\text{eff}})]^{1/2}$, where $\mu_0 M_{\text{eff}}$ is the out-of-plane demagnetization field of NiFe (~0.69 T) and $\gamma$ is the gyromagnetic ratio. The Gilbert damping constant $\alpha$ is ~0.013 as determined from the relation $\Delta = \Delta_0 + 2\pi\alpha f/\gamma$,[33] where $\Delta_0$ is the inhomogeneous linewidth broadening and $\Delta$ is the linewidth. The ST-FMR signal measured at each frequency is superimposed by a symmetric and antisymmetric Lorentzian component, which can be decomposed by fitting $V_{mix}$ to

$$V_{\text{mix}} = V_{\text{sym}} \frac{\Delta^2}{\Delta^2+(B-B_0)^2} + V_{\text{asym}} \frac{\Delta(B-B_0)}{\Delta^2+(B-B_0)^2}, \qquad (1)$$

where $V_{\text{sym}}$ and $V_{\text{asym}}$ are the amplitudes of the symmetric and antisymmetric Lorentzian components, respectively. $V_{\text{sym}}$ and $V_{\text{asym}}$ are related to $\tau_\parallel$ and $\tau_\perp$ by $V_{\text{sym}} = -\frac{I_{rf}}{2}\frac{dR}{d\theta_B}\frac{\cos\theta_B}{\alpha\gamma(2B_0+\mu_0 M_{eff})}\tau_\parallel$ and $V_{\text{asym}} = -\frac{I_{rf}}{2}\frac{dR}{d\theta_B}\frac{\cos\theta_B\sqrt{1+\mu_0 M_{\text{eff}}/B_0}}{\alpha\gamma(2B_0+\mu_0 M_{\text{eff}})}\tau_\perp$,[3-4] where $I_{rf}$ is the rf current flowing through the device and $dR/d\theta_B$ is the angular-dependent magnetoresistance at $\theta_B = 40°$, which is measured separately on the same device.

Figure 2c shows the ST-FMR spectrum measured on a CVD-WTe$_2$ (5 nm)/NiFe (8 nm) device at 6 GHz. In Figure 2c the symmetric ($V_s$) and antisymmetric ($V_a$) Lorentzian components are decomposed from the ST-FMR spectrum. Since the magnitude of $V_s$ (green curve) is the same for positive and negative external magnetic fields, the in-plane spin ($\sigma_\parallel$) induced damping-like torque ($m\times\sigma_\parallel\times m$) is the dominant source of $V_s$. In contrast, we observe that the amplitude of $V_a$ (magenta curve) in the ST-FMR signal is quite different for positive and negative external magnetic fields, which can be attributed to the out-of-plane spin ($\sigma_\perp$) induced damping-like torque ($m\times\sigma_\perp\times m$). As a comparison, we also show the ST-FMR spectra measured on a control sample of Pt (3 nm)/NiFe (6 nm) in Figure 2d with $f$ ranging from 7 to 10 GHz and a rf power of 15 dBm. Similarly, we decompose the ST-FMR signal from Pt/NiFe at 7 GHz to $V_s$ and $V_a$ in Figure 2e. The magnitudes and lineshapes of $V_a$ as well as $V_s$ show the same behavior under both the positive and negative external magnetic fields for the Pt sample, indicating no signature of $\sigma_\perp$. This is consistent with that the two-fold rotational symmetry in Pt/NiFe in which the SOT changes the sign with reversing the external magnetic field.[34]

Next, we quantify the in-plane and out-of-plane spins generated in CVD-WTe$_2$. The charge-to-spin conversion efficiency $\theta_{c\to s}$ that characterizes the strength of the in-plane SOT per unit applied current density at $\theta_B = 0°$ is given by $(2e/\hbar)\sigma_{s,\parallel}/\sigma$,[3] where $\sigma_{s,\parallel}$ and $\sigma$ are the in-plane spin conductivity and the charge conductivity of the SOT material, respectively. The in-plane ($\sigma_{s,\parallel}$) and out-of-plane ($\sigma_{s,\perp}$) spin conductivity are defined as the in-plane and out-of-plane spin-polarized current densities per unit electric field, i.e., $\sigma_{s,\parallel} = J_{s,\parallel}/E = \tau_\parallel M_S t_{FM}/E$ and $\sigma_{s,\perp} = J_{s,\perp}/E = \tau_\perp^{\sigma_{s,\perp}} M_S t_{FM}/E$, where $J_{s,\parallel}$ and $J_{s,\perp}$ are the in-plane and out-of-plane spin-polarized current density absorbed by the ferromagnet at $\theta_B = 0°$, respectively. $E$ is the electric field across the device. $\tau_\perp^{\sigma_{s,\perp}}$ denotes the out-of-plane damping-like torque. To determine $\tau_{\sigma_{s,\perp}}$, we apply the relation $\tau_\perp(-B) = -\tau_\perp^{\sigma_{s,\parallel}+Oe} + \tau_\perp^{\sigma_{s,\perp}}$ and $\tau_\perp(B) = \tau_\perp^{\sigma_{s,\parallel}+Oe} + \tau_\perp^{\sigma_{s,\perp}}$ based on the symmetry argument,[21, 23] where $\tau_\perp^{\sigma_{s,\parallel}+Oe}$ is the sum of field-like torque induced by the Oersted field and the in-plane spin. This analysis method is validated by the angle-dependent ST-FMR measurements (Supporting Information S2).

We can thus determine $\sigma_{s,\|}$ and $\sigma_{s,\perp}$ from $\tau_\|$ and $\tau_\perp^{\sigma_{s,\perp}}$, and they are $(7.36 \pm 0.29)\times10^3$ $(\hbar/2e)$ $(\Omega m)^{-1}$ and $(1.76 \pm 0.18)\times10^3$ $(\hbar/2e)$ $(\Omega m)^{-1}$, respectively, which are values averaged by four devices measured from 6 to 9 GHz (Supporting Information S3). The room temperature resistivity $\rho$ of a 5 nm-CVD-grown WTe$_2$ is ~ 1769.7 $\mu\Omega\cdot$cm, and thus the in-plane and out-of-plane $\theta_{c \to s}$ are determined to be 0.130 and 0.031, respectively, based on $(2e/\hbar)\sigma_{s,\|}\rho$ and $(2e/\hbar)\sigma_{s,\perp}\rho$. More importantly, the signal induced by $\sigma_{s,\perp}$ is not only repeatable in other devices with the same thickness but also in devices with different thicknesses (Supporting Information S3 and S4). It should be noted that using 8-nm NiFe can possibly make our current spin conductivities and charge-to-spin conversion efficiencies underestimated.[35]

In order to understand the microscopic origin of the $\sigma_{s,\|}$ and $\sigma_{s,\perp}$ in CVD-WTe$_2$, the spin Hall conductivity (SHC) is theoretically studied. A 7-monolayer (7L) WTe$_2$ model with a thickness of 4.91 nm is extracted from a bulk Td-WTe$_2$ (space group Pmn2$_1$, #31) Wannier tight-binding Hamiltonian, which reproduces the electronic structures obtained from first-principles calculations. The crystal structure of the orthorhombic Td-phase WTe$_2$ is shown in **Figure 3**a and the band structure of 7L Td-WTe$_2$ is shown in Figure 3b, where the corresponding high symmetry $k$ points are indicated on the corresponding Brillouin zone shown as the inset. The semimetallic feature can be recognized from the overlapping of electron and hole pockets in energy along the $k$ path from Γ to X. For a few-layer WTe$_2$ crystal, two nonsymmorphic symmetries, a two-fold screw rotation and a glide mirror, are absent due to the broken translation symmetry along the $c$-axis (see Figure 3a). The remaining crystal symmetry is a pure mirror reflecting position along the $a$-axis.

Such kind of symmetry reduction has been experimentally shown to harbor unconventional spin Hall effects (SHE). [21-25, 36-37]

CVD-WTe$_2$ may contain domains with different crystal orientation. Thus, the SHE components ($\sigma_{ij}^\alpha$) that can contribute to $\sigma_{s,\parallel}$ and $\sigma_{s,\perp}$ are calculated based on the Kubo formula[38] as shown in Figure 3c and Figure 3d, respectively. The SHE component $\sigma_{ij}^\alpha$ stands for the creation of the spin current $J_{s,i}^\alpha$ flowing along the Cartesian direction $u_i$ carrying spins polarized along $u_\alpha$, under an electric field pointing along $u_j$, i.e., $J_{s,i}^\alpha = \sigma_{ij}^\alpha E_j$. Among the symmetry-allowed SHC components, the terms of $\sigma_{cb}^a$ and $\sigma_{ca}^b$ are the available sources for $\sigma_{s,\parallel}$; however, there is only one term $\sigma_{ca}^c$ that can contribute to $\sigma_{s,\perp}$. The calculated result of $\sigma_{s,\perp}$ is consistent with the previous observation of $\tau_\perp^{\sigma_{s,\perp}}$ in exfoliated WTe$_2$ flakes, where $\tau_\perp^{\sigma_{s,\perp}}$ only presents while the current is applied along the $a$-axis.[21-23] At the chemical potential with charge neutrality ($\mu$=0), $\sigma_{cb}^a$ and $\sigma_{ca}^b$ are 0.7×10$^3$ ($\hbar$/2e) ($\Omega$m)$^{-1}$ and 17.6×10$^3$ ($\hbar$/2e) ($\Omega$m)$^{-1}$, respectively, while $\sigma_{ca}^c$ is 8.9×10$^3$ ($\hbar$/2e) ($\Omega$m)$^{-1}$. It should be pointed out that the as-measured $\sigma_{s,\parallel}$ is larger than $\sigma_{cb}^a$ but smaller than $\sigma_{ca}^b$, which is expected as the grains in the device can be along different directions. Moreover, the as-measured $\sigma_{s,\perp}$ is much smaller than calculated $\sigma_{ca}^c$, which could be due to the cancellation of the out-of-plane spin with opposite direction.[21] The above finding shows that CVD-WTe$_2$ thin films can serve as both in-plane and out-of-plane spin generators.

We further demonstrate current driven magnetization switching using the magneto-optical Kerr effect (MOKE) microscopy in a WTe$_2$ (5 nm)/NiFe (8 nm) heterostructure at room temperature. CVD-WTe$_2$/NiFe is patterned into a rectangular bar with a size of 10 $\mu$m × 25 $\mu$m connected to two electrodes (**Figure 4**a and 4b). The magnetic easy axis of the sample is defined by the shape anisotropy of the NiFe bar and thus along the *y* axis as shown in the hysteresis loops (Figure 4c and 4d). The switching results are shown in Figure 4e and 4f, which are captured by the MOKE microscopy with injecting a current pulse of the width of 30 μs. Before applying the current, the NiFe magnetization (*M*) was first saturated along the +*y* axis by an in-plane external magnetic field. After removal of the external field, a current pulse with the density of $J_c \sim 2.53 \times 10^5$ A cm$^2$ in the WTe$_2$ layer is applied along the +*x* axis. The *M* of the NiFe bar is switched from +*y* to –*y* as indicated by the MOKE image contrast changing from dark (left panel of Figure 4e, denoted with blue arrow on top) to light (right panel of Figure 4e, denoted with red arrow on top). In order to make sure the *M* of NiFe is fully switched from +*y* to –*y*, we apply a large magnetic field (> coercivity) along –*y* after switching, and do not observe any variation of the *M* contrast. We then initialize the *M* of NiFe along the –*y* axis and apply the same amount of current along the –*x* axis. The *M* is switched from –*y* (left panel of Figure 4f, denoted with red arrow on top) to +*y* (right panel of Figure 4f, denoted with blue arrow on top). This bipolar switching results demonstrate a dependence on the current polarity. Note that the heating effect reduces the coercive field of the device so that the spin-orbit torque and the Oersted field can move the domain walls, leading to magnetization switching. The switching behavior is reproducible in a different device (Supporting Information S8).

## 3. Conclusion

In summary, we have investigated the charge-to-spin conversion in CVD-grown WTe$_2$ thin films at room temperature. Both in-plane as well as out-of-plane spin conductivities are observed in CVD-grown WTe$_2$ for the first time. We have further discussed the in-plane and out-of-plane spin polarization in WTe$_2$ based on the first-principles calculations. In addition, we have demonstrated the current driven magnetization switching in a CVD-WTe$_2$/NiFe bilayer. The realization of the out-of-plane polarized spins and the demonstration of magnetization switching based on wafer-scale CVD-WTe$_2$ thin films may provide an impetus for spintronic applications utilizing type-II Weyl semimetals.

## 4. Experimental Section

*Fabrication of ST-FMR devices*: The ST-FMR device shown in the middle panel of Figure 2a was prepared through photolithography and lift-off processes. First, an 8-nm NiFe thin film and 5-nm $SiO_2$ capping layer were deposited on top of $WTe_2$ thin films using magnetron sputtering with a base pressure $< 2\times10^{-9}$ Torr. The deposition rate of NiFe was kept less than 1 nm/min. Then, the $WTe_2$/NiFe bilayer was patterned into strips with a length of 10–25 $\mu$m and width of 15–30 $\mu$m as current channels. Next, a Ta (2 nm)/Cu (150 nm)/Pt (3 nm)-structured waveguide was fabricated.

**Acknowledgements**

This work was supported by SpOT-LITE program (A*STAR grant, A18A6b0057) through RIE2020 funds, Singapore Ministry of Education (MOE) Tier 1 (R263-000-D61-114), Samsung Electronics' University R&D program (Exotic SOT materials/SOT characterization), Singapore National Research Foundation for funding the research under medium-sized centre programme, National Basic Research Program of China (No. 2015CB258400), National Key Research and Development Program of China (No. 2016YFB0700702), and National Natural Science Foundation of China (No. 51402118, 61674063). G.E. acknowledges support from the MOE, Singapore, under AcRF Tier 3 (MOE2018-T3-1-005). C-H.H. and G.L. are supported by MOE2017-T2-1-114.

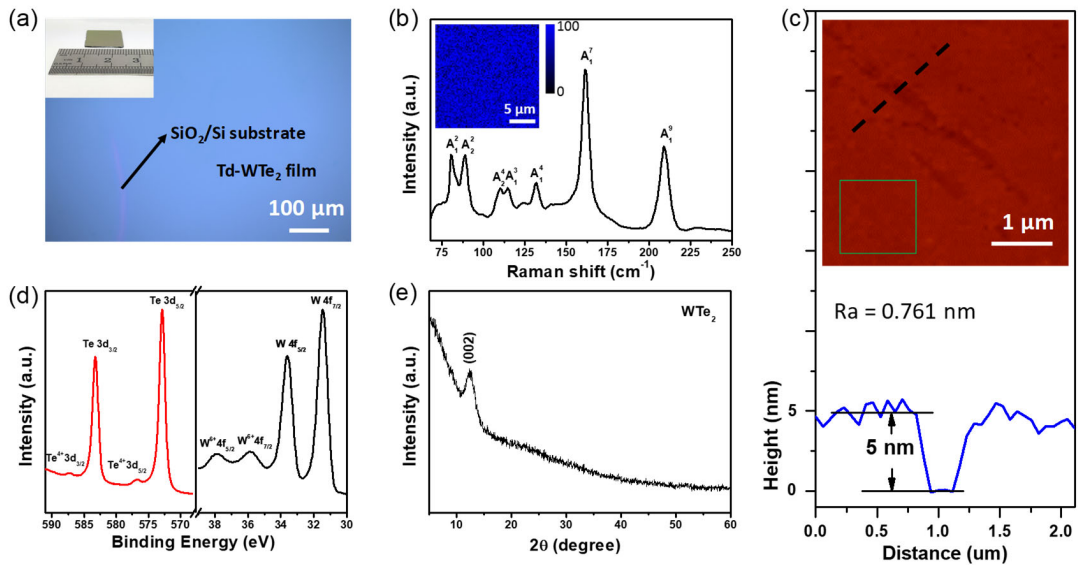

**Figure 1.** a) Optical microscopy image of a typical Td-WTe$_2$ thin film. Inset: a typical 5 nm CVD-grown Td-WTe$_2$ thin film on a Si/SiO$_2$ wafer. b) Raman spectrum for a typical CVD-grown Td-WTe$_2$ (5 nm) thin film on Si/SiO$_2$. Inset: $A_1^7$ Raman mapping of a CVD-grown Td-WTe$_2$ thin film on Si/SiO$_2$. c) Upper: AFM image of a CVD-grown Td-WTe$_2$ thin film on a Si/SiO$_2$ wafer. Lower: AFM height profile along the black dotted line indicated in the AFM image. The green box is the selected area for Ra roughness evaluation (Ra = 0.761 nm). d) XPS spectrum of W 4$f$ and Te 3$d$ electrons. e) XRD spectrum measured from a CVD-grown Td-WTe$_2$ thin film.

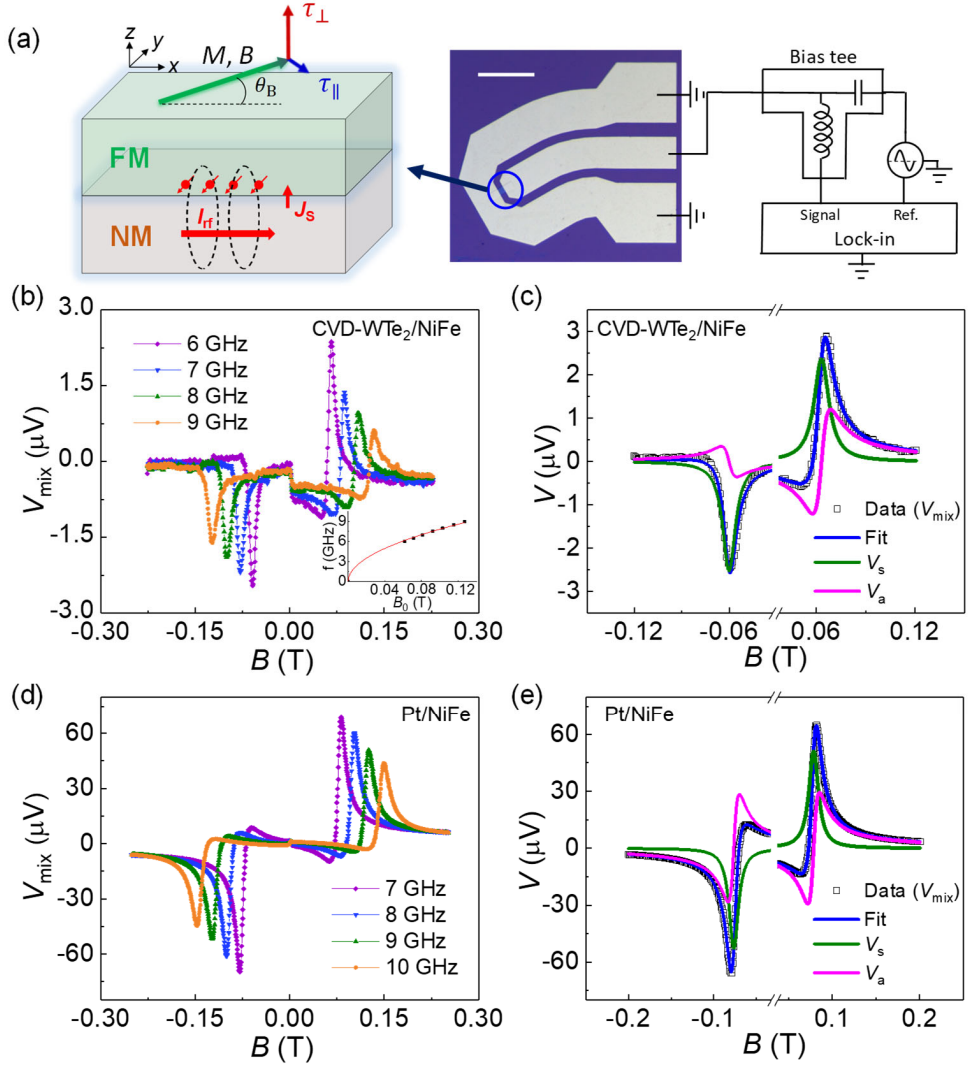

**Figure 2.** a) (left) Schematic diagram of the ST-FMR layer structure, and (right) the measurement circuit connected to a ST-FMR device. The scale bar represents 200 $\mu$m. b) ST-FMR spectra from a CVD-WTe$_2$ (5 nm)/NiFe (8 nm) sample with rf frequencies ranging from 6 to 9 GHz. c) ST-FMR data $V_{mix}$ from a CVD-WTe$_2$ (5 nm)/NiFe (8 nm) sample at a frequency of 6 GHz fitted to Eq. (1). The solid lines are fitting results for the symmetric ($V_s$) (green line) and antisymmetric ($V_a$) (magenta line) Lorentzian components. d) ST-FMR spectra from a Pt (3 nm)/NiFe (6 nm) sample with rf frequencies ranging from 7 to 10 GHz. e) $V_{mix}$ from a Pt (3 nm)/NiFe (6 nm) sample at a frequency of 7 GHz fitted to Eq. (1). The solid lines are fitting results for $V_s$ (green line) and $V_a$ (magenta line) Lorentzian components.

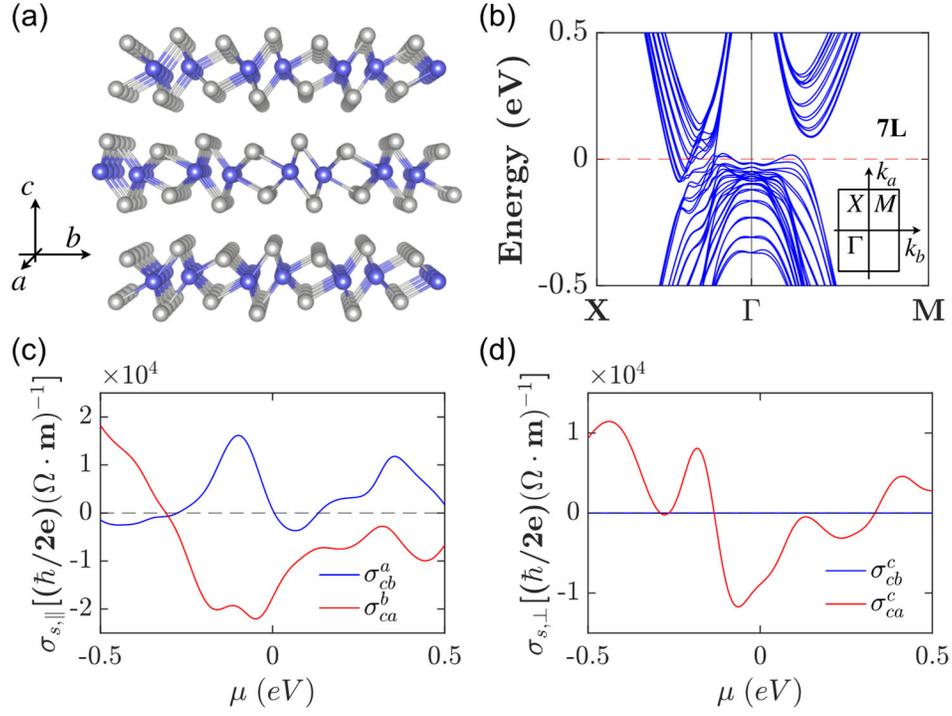

**Figure 3.** Electronic structure and spin Hall conductivity of a WTe$_2$ slab. a) Crystal structure of the orthorhombic Td-phase WTe$_2$. W: blue, Te: gray. b) Band structure of a WTe$_2$ slab with 7 monolayers. c,d) In-plane spin Hall conductivity $\sigma_{s,\parallel}$ (c), and Out-of-plane spin Hall conductivity $\sigma_{s,\perp}$ (d) calculated from a 7 monolayer-WTe$_2$ slab when the electric field is along the *a*-axis (red curves) and *b*-axis (blue curves).

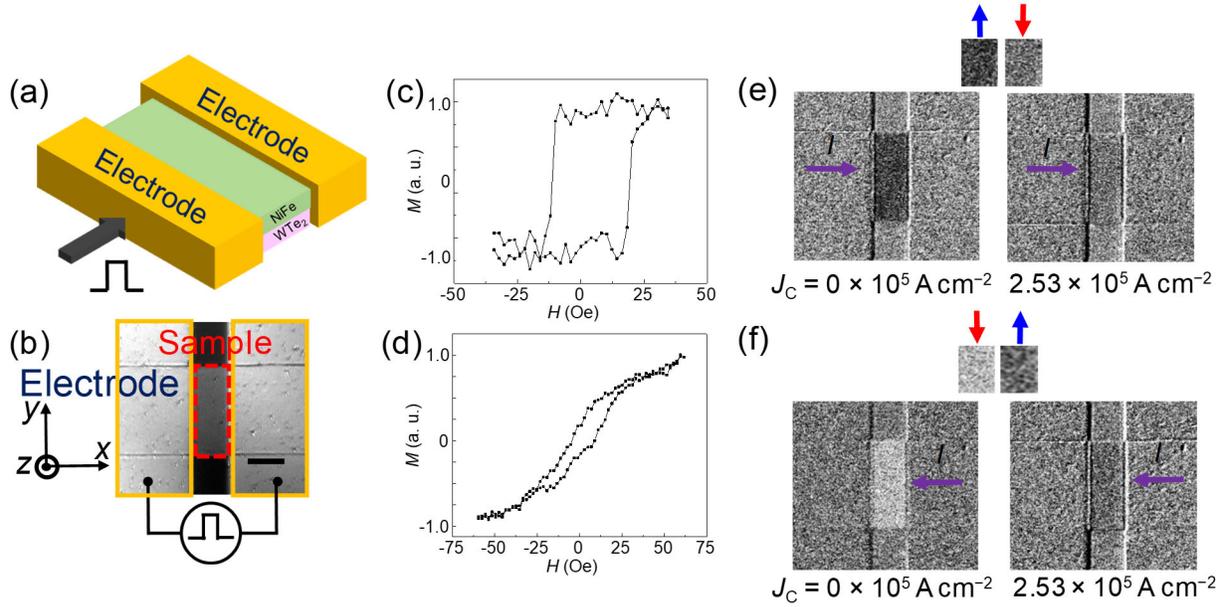

**Figure 4.** a) Schematic of the magnetization switching device shown with measurement configuration. b) Switching device image with electrodes (yellow box). The WTe₂/NiFe region is indicated by the red box. The scale bar represents 10 $\mu$m. c,d) Magnetic hysteresis loops obtained from MOKE by applying an external magnetic field $H$ along the easy $y$-axis (c) and hard $x$-axis (d) in arbitrary units (a.u.). e,f) MOKE images of NiFe before (left) and after (right) applying a pulsed DC current $I$ (indicated by the purple arrow) along the $x$-axis. (e) Magnetization of NiFe is switched from $+y$ (relatively dark contrast, blue arrow) to $-y$ (relatively light contrast, red arrow) by a current pulse $I$ along $+x$. (f) Magnetization of NiFe is switched from $-y$ (relatively light contrast, red arrow) to $+y$ (relatively dark contrast, blue arrow) by a current pulse $I$ along $-x$.